\documentclass[conference]{IEEEtran}
\IEEEoverridecommandlockouts
\usepackage{amsmath,amssymb,amsfonts}
\usepackage{graphicx}
\usepackage{hyperref}
\begin{document}

\title{On the development of an application for the compilation of global sea level changes}

\author{\IEEEauthorblockN{Mihir Odhavji}
\IEEEauthorblockA{\small  \\
\\
Instituto Superior Técnico \\
Universidade de Lisboa, Portugal\\
{\small 	mihir.odhavji@tecnico.ulisboa.pt}
}
\and
\IEEEauthorblockN{Maria Alexandra Oliveira}
\IEEEauthorblockA{\small Centre for Ecology, Evolution and Environmental Changes \\
CHANGE - Institute for Global Change and Sustainability, \\
Faculdade de Ciências\\
Universidade de Lisboa, Portugal \\
{\small maoliveira@ciencias.ulisboa.pt}
}

\and
\IEEEauthorblockN{João Nuno Silva}
\IEEEauthorblockA{\small INESC-ID Lisboa\\
 \\
Instituto Superior Técnico \\
Universidade de Lisboa, Portugal\\
{\small joao.n.silva@inesc-id.pt}
}
}

\maketitle

\begin{abstract}

There is a lot of data about mean sea level variation from studies conducted around the globe. 
This data is dispersed, lacks organization along with standardization, and in most cases, it is not available online.
In some instances, when it is available, it is often in unpractical ways and different formats. 
Analyzing it would be inefficient and very time-consuming. In addition to all of that, to successfully process spatial-temporal data, the user has to be equipped with particular skills and tools used for geographic data like PostGIS, PostgreSQL and GeoAlchemy. 
The presented solution is to develop a web application that solves some of the issues faced by researchers. 
The web application allows the user to add data, be it through forms in a browser or automated with the help of an API. 
The application also assists with data querying, processing and visualization by making tables, showing maps and drawing graphs.
Comparing data points from different areas and publications is also made possible. 
The implemented web application permits the query and storage of spatial-temporal data about mean sea level variation in a simplified, easily accessible and user-friendly manner. 
It will also allow the realization of more global studies.

\end{abstract}


\section{introduction}

Climate change, which can be defined as a modification of climate patterns, has caused worldwide effects owing to changes in weather and climate ex-
tremes, a rise in global surface temperature, glacier melting and shrinking along with worldwide mean
sea level increase\cite{p39}. On the social sphere, these changes are a contributing factor to worsening inequality, economic losses due to a higher number
of natural disasters, crop death and failure related
to drought or out of season rains, which in turn
inflate the food prices\cite{p13}. A rise in sea level is
especially impacting due to its effects on coastal
zones, where big and important cities are mostly
located\cite{p31}. Nonetheless, sea level integrates the effects of diverse complex geological and climatologi-
cal phenomena\cite{p12}. As an example, locally obtained
sea level changes measurements can result from vertical land motion owning to either glacial isostatic
adjustments or tectonics (for example, local glacier
retreat generates a decreased ice burden - or weight
- over landmasses, which in turn generates land uplift), which do not indicate an increase in absolute
mean sea level\cite{p17}. Understanding the climate response at global, regional and local scales from various lines of evidence is crucial to estimate climate related risks and adaptation\cite{p39}. Even though sea
level change represents one of the longest records
using human instruments, going back 300 years in
Europe, these are not enough to describe the non anthropogenic related sea level increase of approx-
imately 120m, which has occurred since the last
glacial maximum (\~19 000 years)\cite{p30, p17}. Therefore, the study of sea level changes is based on geological proxies that show a past position of sea
level. Sea-level index indicators include instrumental evidence (for example from tidal gauges), ancient shorelines (such as raised beaches), biological indicators (such as bivalve shells), intertidal deposits (such as coastal marshes), and archaeological
evidence (for example the position of ancient harbor and port facilities)\cite{p16}.

With an increasing number of studies conducted
and articles published, the amount of data available
about sea level variation is also growing. However,
data was saved on physical paper, which is easily
corruptible, difficult and slow to share at a global
level, can amount to huge stacks very quickly leading places to run out of space. Nonetheless, with
computers becoming more mainstream, some of the
problems were solved with excel sheets. However,
excel comes with its problems like not being able to
create maps and difficulty with collaboration. Every time someone wants to participate, a link has
to be shared and they have to be added to the excel sheet and that is not scalable. The file could
also be made public to solve the scalability issues,
but that is not secure enough. Another argument
against excel is the fact that it is not a relational
database. If everything is on the same sheet, it can
become messy very quickly, if the data is in different sheets then, firstly it can be easy to lose track
of the parameters, secondly, it can not have relations between tables, and thirdly queries cannot be
made. A further problem is the fact that even assuming such a database exists, it is not online. By
having it public the database can increase in size
in a manner that a private one could never. Also,
more people visualizing the Database means more
people looking for errors and easily correcting them
if needed. The problem is that there isn’t a single
database that combines all the data from all the
studies conducted.

The objective is to develop a web tool that aggregates the maximum number of data points taken
from local researchers about the variation of the average sea levels and helps the study of changes in
mean sea level by using data from the holes dug. A
tool that lets researchers collaborate without creating unnecessary friction (such as sharing links and
files). An instrument that shows and creates dynamic maps and graphs, allows for some data query
and exploration along with letting users compare
the variation of the average sea level from different
parts of the world. The tool should also facilitate
the data entries made by the researchers, help them
collaborate more smoothly and do some of the data
standardization required for a better comprehension of the data available.

The expected results from this thesis are the creation of a web application that uses a relational
database, a georeferenced database and allows for
the creation of dynamic maps and graphs. The relational database saves observations that are represented by their geographic coordinates in a georeferenced database, the age, the area, the title of
the paper on which they were published and the
sea level. Using these data points, the application
creates maps and graphs where it is possible to compare different places for the variation of the mean
sea level.

\section{Background}

\subsection{Building and Interpreting Sea levelCurves}

Sea level index points represent the position of past
sea levels and mainly comprise biological, sedimentological, morphological, or archaeological evidence of past sea levels. Each type of sea level indicator is associated with a precision (assumed as an error interval) 
in its relationship with the mean sea level. For example, specific foraminiferal assemblages (foraminifera are single cell organisms
with particular external shells) are nowadays found
in high marsh environments, living only between
the highest astronomical tide and the mean high
water-level \cite{p10}. So, to a great degree, by finding
these shells at lower depths when studying sediment cores, researchers find evidence of shallow water conditions, within a specific height range relative to the existing mean sea level. The next step
is to estimate the age of the sediment and/or shells
with the appropriate age estimation methods, such
as radiocarbon dating, and a local sea level index
point can be established.

Several ages estimation methods are often used in
the establishment of sea level index points, largely
depending on the indicator and time frame under
analysis. Each method has its precision, mainly related to the statistical uncertainty obtained from
the physical or chemical analysis used in the age
determination \cite{p37,p20}. Due to this reason, age estimation of either sediment, biological, or rock samples, is always accompanied by an error.

Sea level curves are normally represented by scatter plots showing the relative sea level plotted
against age, also named sea level index points, at
a local scale. The consideration of precision/errors
associated with the selected sea level indicator, and
region (due to different tidal ranges, for example),
combined with age estimation uncertainties, which
depend on the technique used, are extremely important when comparing sea level index points from
different locations or ages, and are commonly represented by either polygons or error bars \cite{p37,p18}.

Sea level curves can differ considerably in different locations, as they represent the local relative sea level (height of the sea surface relative to
the earth surface), mostly due to also local vertical land movement ]cite{p10} This land movement
is related to several regional processes, including
present and past changes and land ice mass, with
major contributions from surface loading response
to past glacial-isostatic adjustments, plate tectonics, as well as local processes, such as sediment compaction and past tidal range changes \cite{p10}. In fact,
Holocene sea level databases are frequently used
to constrain parameters in Glacial isostatic adjustment models, with important applications in the
current understanding of sea level rise and its geographical variability \cite{p10,p17}. Quantification of relative and absolute sea level changes, the latter also
named global mean sea level rise, is in continuous
development due to the increasing amount of sea
level data, and vertical land movements measured
using geophysics and geodesy equipment \cite{p32}. Taking all these complex inter-related processes, the
development of standardized Holocene sea level and
vertical land-movement databases is pivotal to establish the global mean sea level rise and to have a
better understanding of its changes and relationship
with anthropogenic forcing.

\subsection{Data Representation Libraries}

The web application that is being planned to be
developed has a server side and a client side. For
the client side, HTML\cite{p38} is being considered, as it
is easy to understand, it has a lot of documentation,
forums and tutorials online, it can be embedded
with other languages such as JavaScript\cite{p6} and with
some basic knowledge, is fairly easy to create forms
and tables for data collection and display. Other
than forms and tables, the application should show
personalized maps and plot graphs. However is not
possible with HTML and JavaScript alone to show
personalized maps and plot graphs, so alternatives
were contemplated.

\begin{figure} [htb]
\centering
\begin{minipage}{0.49\columnwidth}
\centering
  \includegraphics[width=\textwidth]{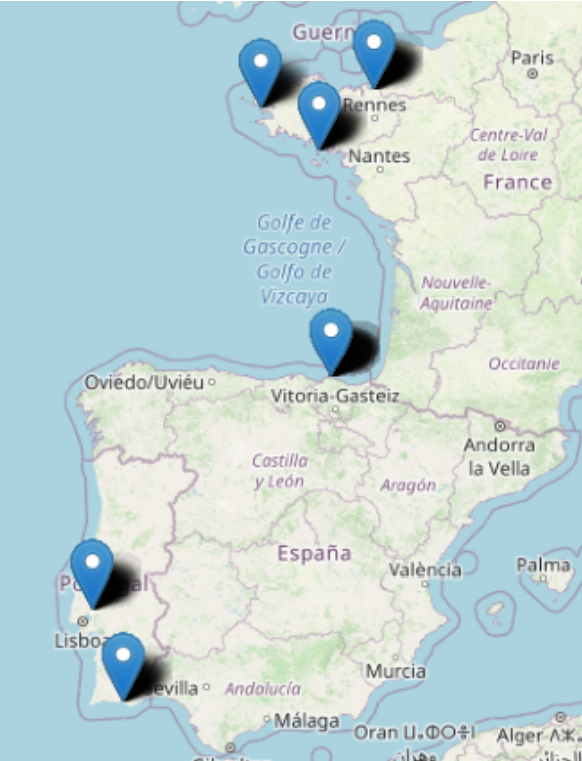}
  
  (a)
\end{minipage}
\begin{minipage}{0.49\columnwidth}
\centering
  \includegraphics[width=\textwidth]{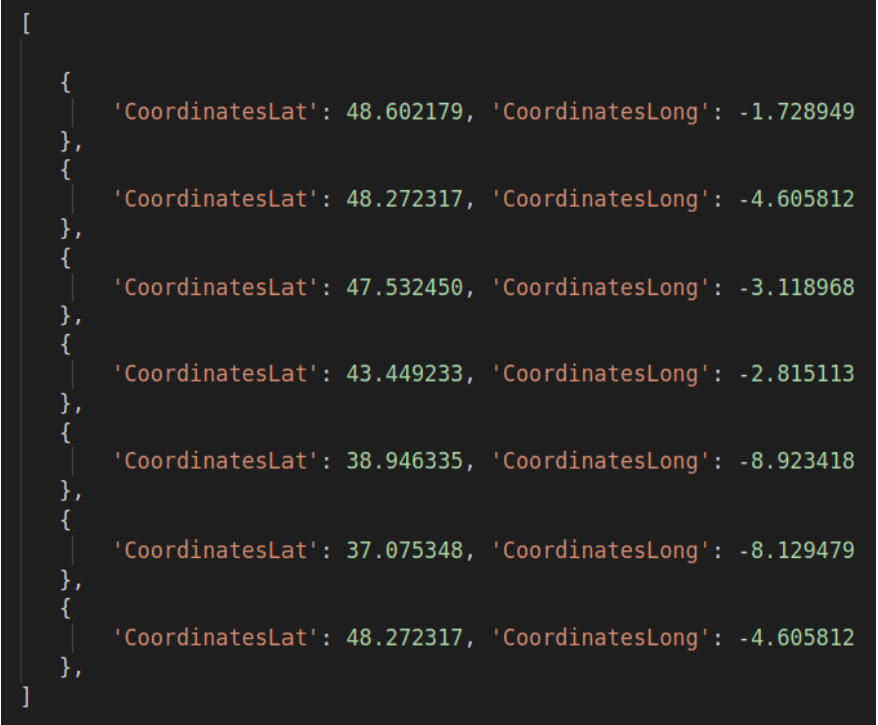}
  
  (b)
\end{minipage}
    
    \caption{Example of a map with coordinates of observations points (a) and the code to do represent them (b)}
    \label{fig:1}
\end{figure}{}

One of the requirements for the application is 
to show maps from a specific area or publication.
These maps shown should have a pin for each of
the observations on that specific area or publication.
 The coordinates of every observation are sent to the browser via a REST/AJAX\cite{p9,p22} request.Considering all the prerequisites, two alternatives
 were shortlisted: Google Maps\cite{p11} and Leaflet\cite{p36}.
 In figure\ref{fig:1}, an example of a map shown to the user
 is provided, along with the JSON\cite{p15} received by
 the browser. The library, in this case, should use
 the data obtained to pin the points on the map.

\begin{figure} [htb]
\centering
\begin{minipage}{0.49\columnwidth}
\centering
  \includegraphics[width=\textwidth]{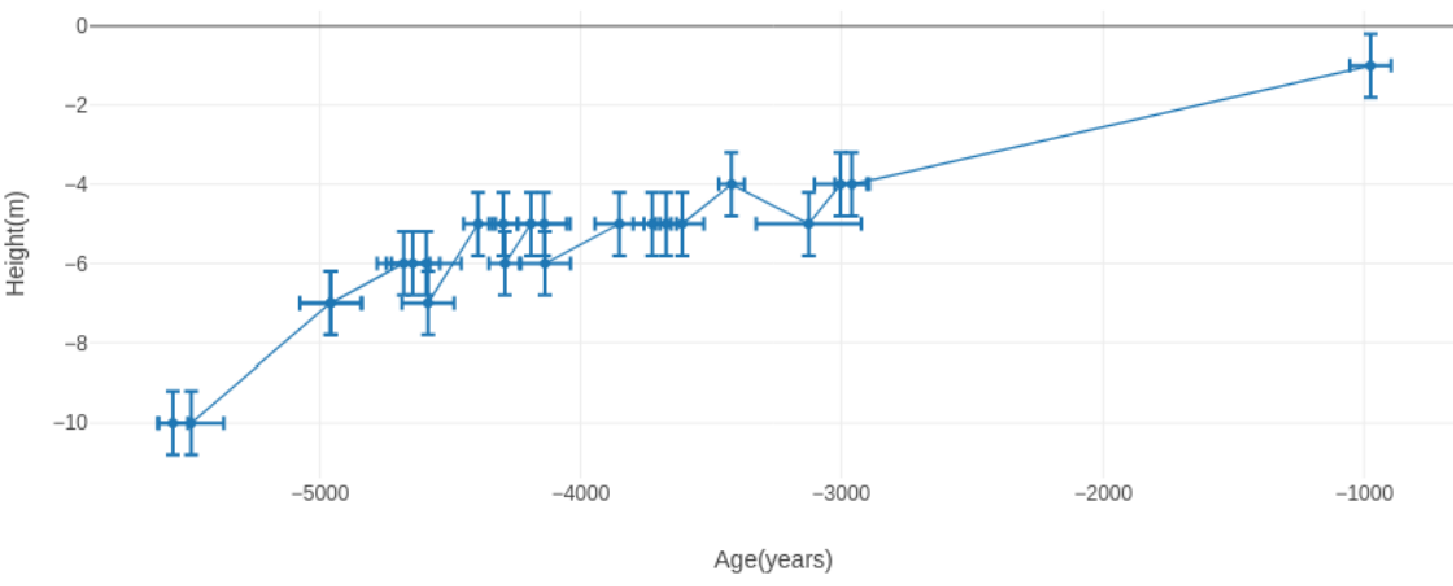}
  
  (a)
\end{minipage}
\begin{minipage}{0.49\columnwidth}
\centering
  \includegraphics[width=\textwidth]{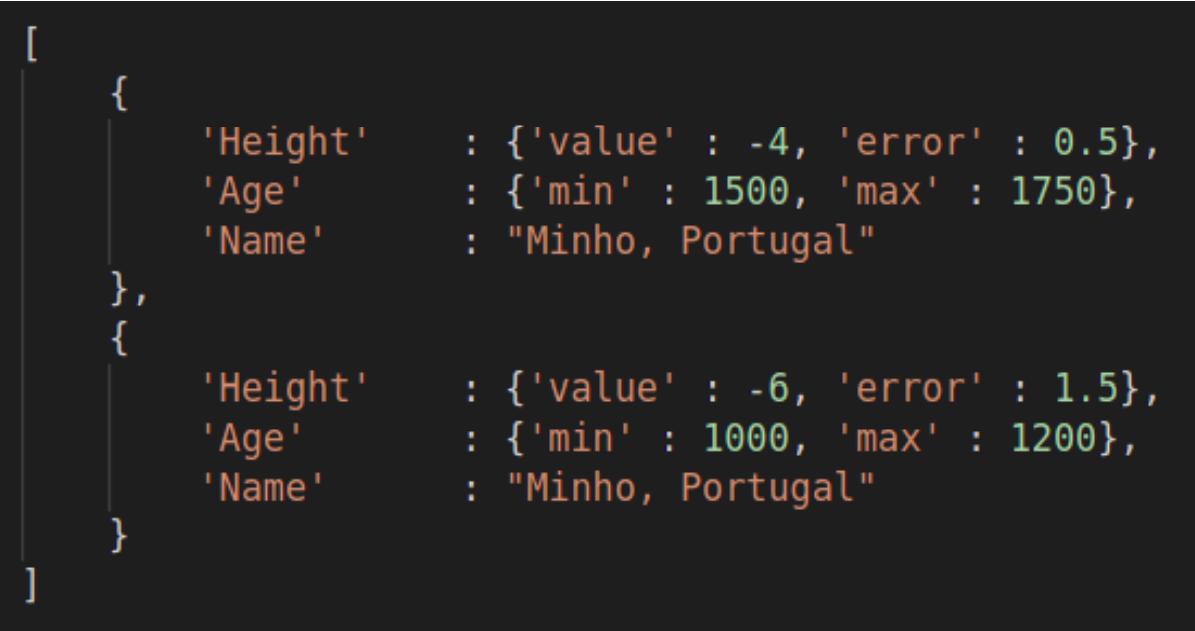}
  
  (b)
\end{minipage}
    
    \caption{Example of a graph plotted with error bars  (a) and the code to do represent them (b)}
    \label{fig:2}
\end{figure}

Another requirement for the application was to
plot graphs for the sea level variation for a specific
area and have the option to compare two or more
areas. With these conditions in mind, two alternatives
 were finalized: Chart.js\cite{p4} and Plotly\cite{p26}. In
 figure\ref{fig:2}, an example of a graph shown to the user
 is provided, along with the JSON\cite{p15} received by
 the browser. The library, in this case, utilizes the
 data received to draw the graphs and the respective
 vertical and horizontal error bars.

{}

\subsection{Geographical Databases}

Being a project closely related to geology and geography, 
it was decided to look for tools that combine
these fields with servers and Databases. ArcGIS\cite{p8},
GeoServer\cite{p21} MySQL\cite{p23} and PostgreSQL\cite{p35}
were found. ArcGIS is a platform to generate, organize, 
distribute and inspect spatial data. It can
be run locally or on cloud services. GeoServer is
a combination of a Web server with a Web Gis
software and a database. It is open-source and 
allows users to share, manage and organize geospatial data. MySQL is a relational database management 
system. MySQL spatial extensions permit the users with the creation, storage and inquiry of spatial data through spatial data types and
functions\cite{p24}. PostgreSQL is an open source object--relational database system that is more closely related 
to SQL, however, to use geographical libraries
an extension named PostGIS\cite{p27} has to be added to
the database. A further advantage of PostGIS is its
capacity for doing spatial queries. Using PostGIS
you can create a variety of geographical data types
like Points, Line, Circles, Polygons, etc.\cite{p2}. These
data types can interact with each other to make
queries such as the meeting point of two lines or
the intersection of a line with a shape\cite{p28}. This
functionality is not only limited to geometry 
entities, it can be done with geographical bodies as well,
namely lakes, rivers mountains, etc.

\subsection{ORM}
The browser can create queries for a database that
is connected to the server. A system is essential
to translate the server’s request to the database.
That is where ORM enters. Object-relational
mapping (ORM)\cite{p1} is a technology that lets the
user add, query, change and delete data from a
database employing an object-oriented paradigm.
The ORM library is an entirely normal library
written in the same language as the rest of the
code. The big advantage of using ORM is that it
encapsulates the code required to manipulate the
data, so SQL is not used. Direct interaction is
done with the objects that use the same language
as you. SQLAlchemy allows the use of various
database management systems with just a configuration change. For the server, Python is used. Having that in mind, options for ORM were searched,
and two of them were shortlisted: SQLAlchemy\cite{p33}
and Django\cite{p5}. The main difference between using SQL and SQLAlchemy is that, while using the
former generic table rows are received, but using
SQLAlchemy a list with objects is returned, which
can be immediately used for further development.

\section{Implementation}

\subsection{Requirements}
Before starting the implementation, it is important
to outline the requirements of the program that will
be developed :

\begin{itemize}
 \item The application should be remotely accessible.
 \item The application should store the data generated by the researcher about proxies for the variation of sea levels (proxy data).
 \item The proxy data about sea level variation should be organized in areas.
 \item The application should store the publications where the data was published.
 \item The application should store information about the indicator used to estimate the age of the sample.
 \item Each data point should include the following information: coordinates, estimated age and height.
 \item The application should present the data in maps (global with all the data and particular to a publication).
 \item The application should create plots with the representation of the data for a specific area. 
 \item The plots should contain error bars for the age and height estimations.
 \item The user should be allowed to compare different areas in the same plot.
 \item The user should be able to download the data in tabular format.
 \item The application should contain local information about the vertical land movement.
 \item The application should provide forms for the introduction of data to the database.
\end{itemize}

\subsection{Data Model}
\begin{figure}[!ht]
        \centering
        \includegraphics[width=\columnwidth]{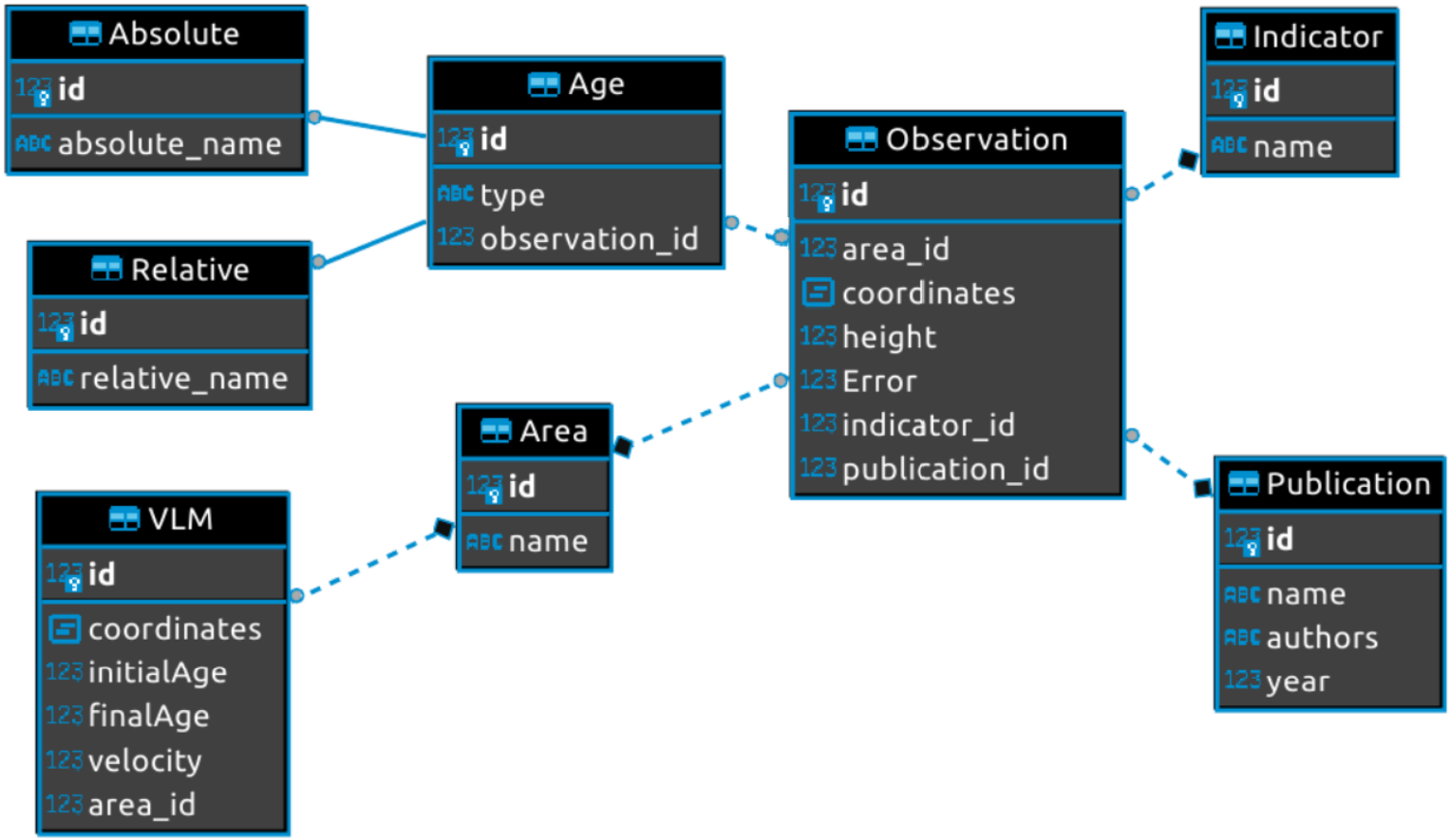}
        \caption{Unified Modelling Language of the data structures.}
        \label{fig:3}
        \end{figure}

In order to solve the problems explained, it was
decided to build a relational database. The final
result is illustrated in figure~\ref{fig:3}.

As seen in figure~\ref{fig:3}, there are many tables that
save different data proxies. Every table has a primary key and its own ID.
The tables are:
\begin{itemize}
\item Area - a table that saves the area name and an ID number that is unique to each area.
\item Indicator - a table that saves the indicator name and an ID number that is unique to each Indicator.
\item Publication - a table that saves the publication name, its authors and the year on which it was published. It also has a unique ID number for each row.
\item Vertical Land Movement or VLM - a table that saves the coordinates of the point, the initial and final ages, the velocity in mm/yr, the ID of the area that is related to and a unique ID for each of its own entries.
\item Observation - a table that saves the coordinates and height of the sample, the error (in meters) associated with the Indicator, the ID of the area and publication and a unique ID for each observation.
\item  Age - a table that saves the ID of the observation that it is related to, the type of age (Absolute or Relative) and a unique ID for each age. The Absolute and Relative ages table inherit the ID from its parent table age\cite{p34}.

\end{itemize}

\subsection{Architecture}

The three main components of the program devel-
oped are the browser, the server and the database.

The browser is used as the client of the server.
It is where the HTML\cite{p38} presents forms to in-
troduce data along with showing tables, graphs
and maps. The browser has the succeeding
modules:
\begin{itemize}
\item Add Data - A module for adding data to the database.
\item View Map - A module for viewing maps with the coordinates sent by the server.
\item Plot Graph - A module for plotting graphs with the information received from the server.
\item Download CSV - It is a module that transforms data listed on the web page to a CSV type file\cite{p3}
\end{itemize}

 The server receives requests sent by the client(browser) and responds to them. 
 To successfully respond to them, it needs to convert the requests to queries for the database. 
 To do that translation, it uses ORM\cite{p1}. 
 After receiving the result from the database, it sends the output to the browser to show it to the user.
The server is comprised of the following modules :

\begin{itemize}
\item Add Data - A module for adding data to the database.
\item List Data - A module to list data from the database
\item Vertical Land Movement - A module that stores the data related to vertical land movement.
\item View Map - A module that plots the observations’ coordinates on a flat 2D map.
\item Plot Graph - A module that draws the graphs for sea level variation throughout the ages.
\end{itemize}

 The database receives queries from the server and it responds to them. It has to correctly save geographical data such as the point coordinates.
 
\subsection{Support System}

To successfully implement this web application, it
was important to plan in which language to code the
three main modules of the program. The browser
is mainly coded in HTML and JavaScript\cite{p6}, the
server in Python\cite{p29}, using the Flask framework\cite{p25},
and the Database in PostgreSQL\cite{p35}.

Starting from the browser, HTML is used to create
 the web pages, to receive input from the users
and to show them the output. JavaScript is utilized
to create requests that are sent to the Flask server.
The requests are read by the HTML from the user,
built by JavaScript and sent to the server with
help of REST and AJAX\cite{p9,p22} Afterward, the
JavaScript receives the response and displays it using
 to the user using HTML. To produce the maps,
JavaScript’s Leaflet\cite{p36} library is used. Leaflet is
used because of how coding friendly and well 
documented it is. For plotting the graphs it is used the
Plotly\cite{p26} library. Plotly’s easiness and lack of friction
 in processes such as adding data to the plot was
the key factor, so it was decided to go ahead with
Plotly. Additional advantages of using Plotly are
the fact that the user can personalize the zoom settings
 as well as download the graph plotted to the
computer directly, without having to take screenshots.

Relative to the server, it was coded in python using
 the Flask framework, mainly due to its simplicity
 and velocity. The server has the responsibility to
receive requests sent by the browser and give them
to the ORM with to query the database. After trying
 Django\cite{p5} and SQLAlchemy\cite{p33}, SQLAlchemy
was elected to be the bridge between server and
database because the author of this thesis has
some experience with it from a previous project.
As the database has geographical elements, the
GEOAlchemy\cite{p7} library is used to add support
for spatial data types. Using SQLAlchemy and
GeoAlchemy libraries, the first is used to translate
the queries received by the server to SQL\cite{p14} and
the latter to assist the first regarding spatial queries.
With GeoAlchemy it is also possible to add certain
geographical data types such as ’POINT’ to represent
 the latitude and longitude of an observation.

At last, the Database selected is PostgreSQL. It
is an advanced form of SQL that allows the use
of the normal SQL functions like query, primary
keys, triggers, etc plus the use of Geospatial tools
like GeoAlchemy and its Geometry library to enable
 the creation of POINT data type. PostGIS
is also installed as an extender for PostgreSQL.
Other options like MySQL\cite{p23} were rejected because
 GeoAlchemy only supports PostGIS, while
GeoServer\cite{p21} and ArcGIS\cite{p8} were very complex to
start programming.

\subsection{REST End Points}

For a secure and trustworthy connection between
browser and server, it was decided to implement it
using REST endpoints.
The REST Endpoints Implemented are the fol-
lowing :
\begin{enumerate}

\item ”/API/Area/”, ’GET’ - Used to get the list of all areas.
\item ”/API/Pub/”, ’GET’ - Utilized to get the list of all publications.
\item ”/API/Obs/”, ’GET’ - It has the objective to get the list of all observations.
\item ”/API/Indicator/”, ’GET’ - Implemented to get the list of all indicators.
\item ”/API/VLM/”, ’GET’ - Exercised to get the list of all vertical land movements.
\item ”/API/GetName/’, ’POST’ - It has the goal to get a name of area or publication.
\item ”/API/GetObservations/”, ’POST’  - Employed to get a list of observations based on ID.
\item ”/API/AddArea/”, ’POST’  - Request used to create a new area.
\item ”/API/AddPub/”, ’POST’ - Request utilized to create a new publication.
\item ”/API/AddObs/”, ’POST’ - Request sent to create a new observation.
\item  ”/API/AddInd/”, ’POST’ - Request used to create a new indicator.
\item ”/API/AddVLM/”, ’POST’ - Request used to create a new vertical land movement.
\item ”/API/GetChart/”, ’POST’ - created to plot a graph.
\end{enumerate}
The returned values are then utilized to create the
output shown to the user on the browser such as
tables, maps and Graphs.

\subsection{Software Organization}

With the browser and database already discussed,
the focus of this section is on the server. The server
is made of smaller modules. These modules are 
Add Data, List Data,  View Map, Plot Graph and Vertical Land Movement.

Add Data is a module that adds data to the database. It can be an area, publication, ob servation, indicator or vertical land movement.
To help with the standardization while adding data, an algorithm was implemented that converts ages from the Before Present Scale to the
AD/BC scale. 
It does so by taking 1950 and subtracting the age introduced in the Before Present Scale. 
It utilizes the following REST endpoints already described in section 3.5:
\begin{itemize}
\item ”/API/AddArea/”, ’POST’
\item ”/API/AddPub/”, ’POST’
\item ”/API/AddObs/”, ’POST’
\item ”/API/AddInd/”, ’POST’
\item ”/API/AddVLM/”, ’POST’

\end{itemize}

List Data is a module used to list data from
the database. It can be an area, publication,
observation, indicator or vertical land movement. There is also the possibility to list observations for a particular area or publication.
The module makes use of the upcoming REST
endpoints already explained in section 3.5:
\begin{itemize}
\item ”/API/Area/”, ’GET’
\item ”/API/Pub/”, ’GET’
\item ”/API/Obs/”, ’GET’
\item ”/API/Indicator/”, ’GET’
\item ”/API/VLM/”, ’GET’
\item ”/API/GetName/’, ’POST’
\item ”/API/GetObservations/”, ’POST’
\end{itemize}

View Map is the module that plots the observations’ coordinates on a flat 2D map. Users can see all the points on the Database or filter
for a particular area or publication. It uses the ensuing REST endpoints described in section
3.5:
\begin{itemize}
\item ”/API/Obs/”, ’GET’
\item ”/API/GetObservations/”, ’POST’

\end{itemize}

Plot Graph is a module that draws the graphs for sea level variation throughout the
ages. You can see it for each area individually,
see the graphs for all the areas simultaneously
and able or disable certain areas. The module also draws the horizontal and vertical error
bars for the graphs. It makes use of the follow-
ing REST endpoint described in section 3.5:
\begin{itemize}
\item ”/API/GetChart/”, ’POST’
\end{itemize} 

Vertical Land Movement is the module
that stores data related to the vertical land
movement and later can be used for rectifying
the data introduced on the database by applying a correcting factor related to tectonic movement and geographic location. It does not use
any of the REST endpoints explained in section
3.5.

\section{Results}

This section presents how the requirements were fulfilled in the project.

\subsection{The application should be remotely accessible}
The creation of a web application satisfies this requirement. The historical sea level application runs
on a browser, it is connected to a server that communicates to a database using an ORM. An API
is also provided to send requests to the server and
received information. It is also possible to develop
other web applications that connect to the server
and show data differently.

\subsection{The application should store the data generated by the researcher about proxies for the variation of sea levels (proxy data)}
The creation of a database with the model given in
figure~\ref{fig:3}, utilized to store the different data types re-
lated to an observation meets this requirement. The
data is not only stored on a georeferenced database
but there is also the possibility of doing spatial
queries. This can be a foundation, where future
works could be built.

\subsection{The proxy data about sea level variation should be organized in areas}
According to the UML provided in figure~\ref{fig:3}, a table
with the name Area is created, that saves the name
of the area and a unique ID to identify it. This
table is related to the Observation table as well as
the Vertical Land Movement table. Therefore, the proxy data is organized in areas. The papers, from where the data points are taken, organize them in
areas as well. Those areas are the ones that later are plotted in the graphs.

\subsection{The application should store the publications where the data was published}
As stated by the UML given in figure~\ref{fig:3}, a table with the name Publication is created, that saves the name of the publication, the name of its’ authors, the year it was published and possesses a unique ID to identify it.

\subsection{The application should store information about the indicator used to estimate the age of the sample}

According to the UML given in figure~\ref{fig:3}, a table with the name Indicator is created, that saves the
name of the indicator along with a unique ID to identify it. This table is related to the Observation table as well.

\subsection{ Each data point should include the following information: coordinates, estimated age and height}
As stated by the UML provided in figure~\ref{fig:3}, a table
with the name Observation is created, that saves
the coordinates and the height they were found. It
is related to the Age table, which saves the Age
type(Absolute or Relative) as well as the value.

\subsection{The application should present the datain maps (global with all the data and particular to a publication)}
The application can show a list of all the publications in the database. By selecting one, it is possible
to see a map with the coordinates points of observations taken from that publication. The application
also has a tab called Map to see the location of all observations on the database.

\subsection{The application should create plots with the representation of the data for a specific area}
\begin{figure}[!ht]
        \centering
        \includegraphics[width=\columnwidth]{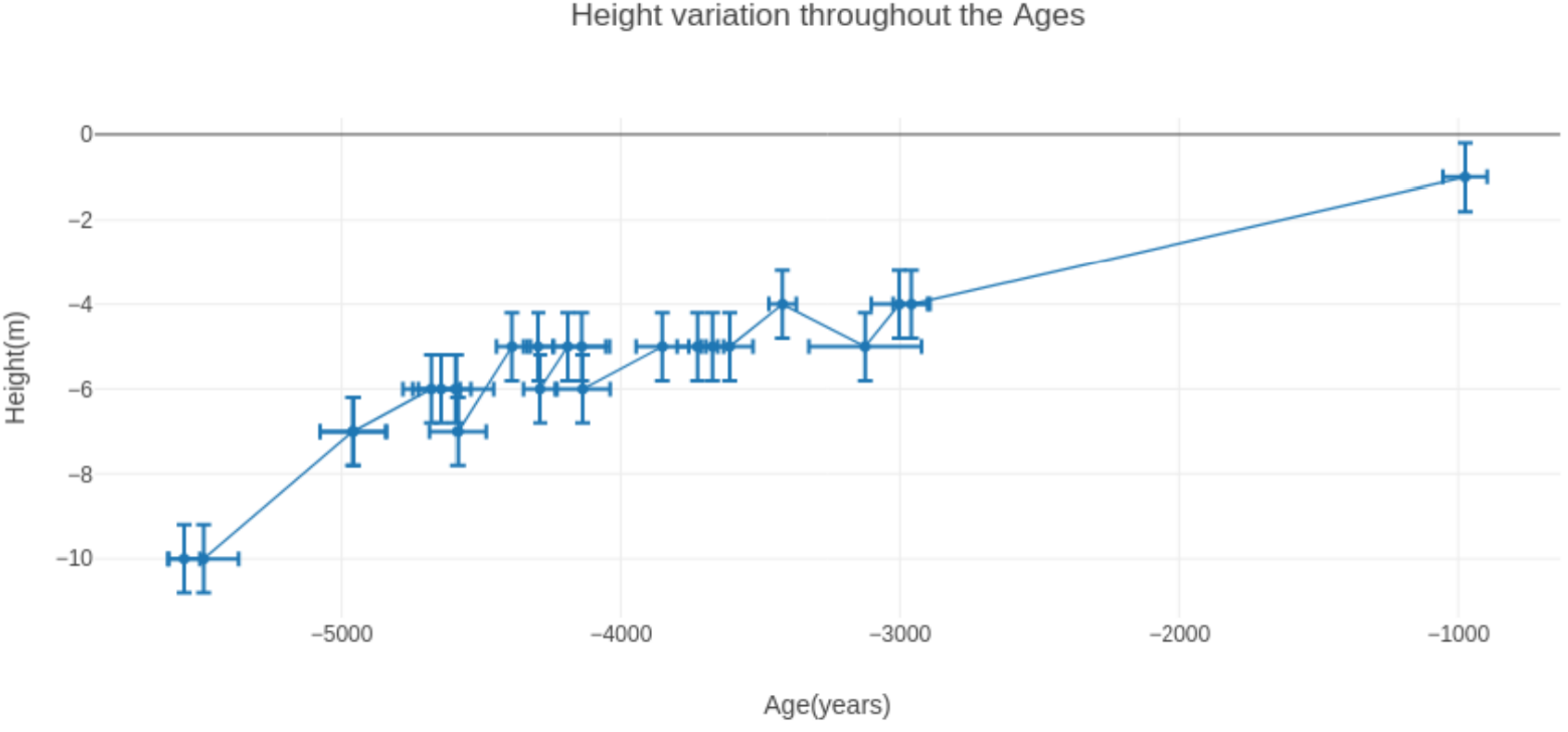}
        \caption{Graph plotted with data from only one area.}
        \label{fig:4}
        \end{figure}

The application gives the possibility to see a list ofall areas in the database. By selecting one, it isfeasible to spot a graph with the data point aboutthat area. An example is given in figure~\ref{fig:4}.

\subsection{The plots should contain error bars for the age and height estimations}
As seen in figure 4, the plots have error bars. The
vertical error bars are related to the height estimations, while the horizontal bars are related to
the age estimation. Compared with the graph from
cite{p19}, it can be concluded that the graphs plotted by
the application are similar to the graphs drawn on
the research papers.

\subsection{The user should be allowed to compare different areas in the same plot}
The  application  has the option to see the graphs of
all areas in the same plot. On that plot, there is a
choice to hide and unhide certain areas.

\subsection{The  user should be able to download the data in tabular format}
To download the data from the database to your
device, at the end of every list on the web application a download button is placed. By clicking on it,
the data from the table that the user is consulting
gets written to a CSV file\cite{3} and downloaded.
As an example, if the table with the list of observations for a particular area was to be transformed
and downloaded, the resulting CSV file would have
the following columns : ID, Coordinates, Height,
Age, Indicator and Error. The data in the down-
loaded CSV file can be processed and used with
other tools or applications.

\subsection{The application should contain local information about the vertical land movement}
As claimed by the UML provided in figure 3, a table
with the name Vertical Land Movement is created,
that saves the coordinates of the point, the initial
and final Ages, the velocity and has a unique ID
to identify it. This table is related to the area ta-
ble as well. Thus, the application contains local
information about vertical land movement. Besides
the fact that data about vertical land movement is
only stored and not processed, its’ existence in the
database allows for the further development and implementation of a simpler model. This tool will be
useful for future work about the relation between
vertical land movement and mean sea level variation because it is possible to realize studies at a
small and local scale along with at a larger and
global scale.

\subsection{The application should provide forms for the introduction of data to the database}
The application provides forms for the user to in-
troduce data into the database. The form seen in
figure 5 is for adding observation to the database.
Similar forms are available to introduce the differ-
ent data types: Area, Publication, Indicator and
Vertical Land Movement.

        \begin{figure}[!ht]
        \centering
        \includegraphics[width=\columnwidth]{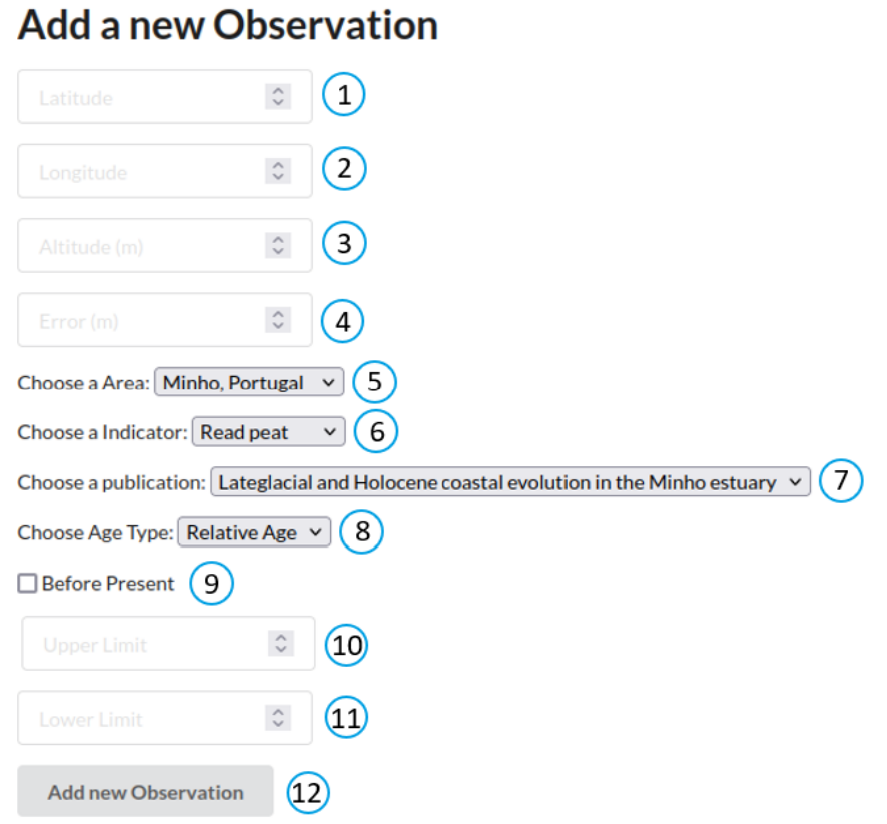}
        \caption{Form utilized to introduce an observation into the database.}
        \label{fig:5}
        \end{figure}
      
The input numbers in the figure 5 are:
\begin{enumerate}
\item To add the Latitude coordinate. It has to
be between -90 and 90 degrees.
\item To put the Longitude coordinate. It has to
be between -180 and 180 degrees.
\item To introduce the altitude value.
\item To write the error value(in meters).
\item To choose the area related to the observation.
\item To select the indicator used to estimate the observation’s age.
\item To pick the publication on which the observation was published.
\item For choosing the age type - absolute or relative. If absolute is selected, the text box on number eleven is hidden, and only the text box on number ten is used to input the age value.
\item If the Before Present option is selected, it automatically converts the Age values to theAD/BC scale.
\item To input the Upper Limit in case the Agetype is relative, in case the age type is absolute,this is the only text box used to input the value.
\item To input the Lower Limit in case the Agetype is relative, if it is absolute, then this textbox is hidden.
\item Button used to submit data into the database.

\end{enumerate}

To ease the navigation between pages of the web
application, a menu was created that is visible on
all the pages. The menu has seven options :
\begin{itemize}
\item Home - Brings the user back to the homepage.
\item Area - To add or list all areas.
\item Publication - To add or list all publications.
\item Observation - To add or list all observations.
Indicator - To add or list all indicators.
\item Map - To see a map with all the different
coordinates of the observations stored in the
database.
\item Graphs - To see a graph with the mean sea
level variation of all the areas stored in the
database.
\end{itemize}

While hovering above some of the options, a drop
down menu appears with two options: add or list
all.

 Conclusions
Until now, there was not a platform that aggregated
spatial information and allowed at a global scale the
processing of data related to mean sea level variation and vertical land movement. With this thesis, a web application was created that aggregates
data from numerous publications and makes them
available in an organized and researcher friendly
manner. The application can be accessed by anyone from a browser. A REST API was also developed, so requests can also be automated. The user
can also add data to the database in a very simple
and frictionless way. The data added is standard-
ized, especially the age value can be converted from
Before Present to the AD/BC scale. Data about
 the vertical land movement is also stored on the
database and can be used for further work. The
developed application lists the data in a clean and
organized manner. That data can be viewed online or downloaded to a CSV file for other works.
The developed application also has the possibility
of viewing maps and plotting graphs. Additionally,
there is the possibility of data comparison. The
plots about the sea level variation throughout the
ges can have graphs from more than one area enabling, this way, the comparison between two or
more areas. This project, for the first time, allows the creation of interactive graphs (with features like zoom in/out and selecting which areas to
hide or show) about the mean sea level variation.
 This new way of data presentation not only permits,
easy comparison between distinct areas but also for
the availability of this content on the web, to in crease the visibility of such problems. The graphs
plotted are a tool used by specialists, who try to
 understand what kind of events occurred in those
areas. However, these types of graphs help identify the time intervals on which the events might
have happened. The current version only stores the
values related to the vertical land movement. It is
also the first time, that it was possible to aggregate on one platform these types of data at a global
evel. This will allow scientists and data modellers,
the development of new theories and/or models to
explain the differences between the various areas.
Using the spatial queries abilities of PostGIS, combined with the data available on the database about
the vertical land movement, mainly the geographic
coordinates, a model can be developed that automatically suggests corrections based on the location
of the observations.

\bibliographystyle{IEEEtran}
\bibliography{./Bibliography}

% Generated by IEEEtran.bst, version: 1.12 (2007/01/11)
\begin{thebibliography}{10}
\providecommand{\url}[1]{#1}
\csname url@samestyle\endcsname
\providecommand{\newblock}{\relax}
\providecommand{\bibinfo}[2]{#2}
\providecommand{\BIBentrySTDinterwordspacing}{\spaceskip=0pt\relax}
\providecommand{\BIBentryALTinterwordstretchfactor}{4}
\providecommand{\BIBentryALTinterwordspacing}{\spaceskip=\fontdimen2\font plus
\BIBentryALTinterwordstretchfactor\fontdimen3\font minus
  \fontdimen4\font\relax}
\providecommand{\BIBforeignlanguage}[2]{{%
\expandafter\ifx\csname l@#1\endcsname\relax
\typeout{** WARNING: IEEEtran.bst: No hyphenation pattern has been}%
\typeout{** loaded for the language `#1'. Using the pattern for}%
\typeout{** the default language instead.}%
\else
\language=\csname l@#1\endcsname
\fi
#2}}
\providecommand{\BIBdecl}{\relax}
\BIBdecl

\bibitem{p39}
V.~Masson-Delmotte, P.~Zhai, A.~Pirani, S.~L. Connors, C.~P{\'e}an, S.~Berger,
  N.~Caud, Y.~Chen, L.~Goldfarb, M.~Gomis \emph{et~al.}, ``Climate change 2021:
  the physical science basis,'' \emph{Contribution of working group I to the
  sixth assessment report of the intergovernmental panel on climate change},
  vol.~2, 2021.

\bibitem{p13}
S.~Hallegatte, \emph{Shock waves: managing the impacts of climate change on
  poverty}.\hskip 1em plus 0.5em minus 0.4em\relax World Bank Publications,
  2016.

\bibitem{p31}
C.~Small and R.~J. Nicholls, ``A global analysis of human settlement in coastal
  zones,'' \emph{Journal of coastal research}, pp. 584--599, 2003.

\bibitem{p12}
V.~Gornitz, ``Sea level change, post-glacial,'' \emph{Encyclopedia of
  paleoclimatology and ancient environments}, pp. 887--893, 2009.

\bibitem{p17}
\BIBentryALTinterwordspacing
N.~S. Khan, B.~P. Horton, S.~Engelhart, A.~Rovere, M.~Vacchi, E.~L. Ashe, T.~E.
  Törnqvist, A.~Dutton, M.~P. Hijma, and I.~Shennan, ``Inception of a global
  atlas of sea levels since the last glacial maximum,'' \emph{Quaternary
  Science Reviews}, vol. 220, pp. 359--371, 2019. [Online]. Available:
  \url{https://www.sciencedirect.com/science/article/pii/S0277379119306468}
\BIBentrySTDinterwordspacing

\bibitem{p30}
M.~Siddall, E.~J. Rohling, A.~Almogi-Labin, C.~Hemleben, D.~Meischner,
  I.~Schmelzer, and D.~Smeed, ``Sea-level fluctuations during the last glacial
  cycle,'' \emph{Nature}, vol. 423, no. 6942, pp. 853--858, 2003.

\bibitem{p16}
\BIBentryALTinterwordspacing
M.~S. Kearney, \emph{Sea Level Indicators}.\hskip 1em plus 0.5em minus
  0.4em\relax Dordrecht: Springer Netherlands, 2009, pp. 899--902. [Online].
  Available: \url{https://doi.org/10.1007/978-1-4020-4411-3_209}
\BIBentrySTDinterwordspacing

\bibitem{p10}
A.~Garc{\'\i}a-Artola, P.~St{\'e}phan, A.~Cearreta, R.~E. Kopp, N.~S. Khan, and
  B.~P. Horton, ``Holocene sea-level database from the atlantic coast of
  europe,'' \emph{Quaternary Science Reviews}, vol. 196, pp. 177--192, 2018.

\bibitem{p37}
M.~Walker, \emph{Quaternary dating methods}.\hskip 1em plus 0.5em minus
  0.4em\relax John Wiley and Sons, 2005.

\bibitem{p20}
\BIBentryALTinterwordspacing
W.~Li, X.~Li, X.~Mei, F.~Zhang, J.~Xu, C.~Liu, C.~Wei, and Q.~Liu, ``A review
  of current and emerging approaches for quaternary marine sediment dating,''
  \emph{Science of The Total Environment}, vol. 780, p. 146522, 2021. [Online].
  Available:
  \url{https://www.sciencedirect.com/science/article/pii/S0048969721015904}
\BIBentrySTDinterwordspacing

\bibitem{p18}
\BIBentryALTinterwordspacing
K.~Lambeck, C.~D. Woodroffe, F.~Antonioli, M.~Anzidei, W.~R. Gehrels,
  J.~Laborel, and A.~J. Wright, \emph{Paleoenvironmental Records, Geophysical
  Modeling, and Reconstruction of Sea-Level Trends and Variability on
  Centennial and Longer Timescales}.\hskip 1em plus 0.5em minus 0.4em\relax
  John Wiley and Sons, Ltd, 2010, ch.~4, pp. 61--121. [Online]. Available:
  \url{https://onlinelibrary.wiley.com/doi/abs/10.1002/9781444323276.ch4}
\BIBentrySTDinterwordspacing

\bibitem{p32}
G.~Spada, ``Glacial isostatic adjustment and contemporary sea level rise: An
  overview,'' \emph{Integrative Study of the Mean Sea Level and Its
  Components}, pp. 155--187, 2017.

\bibitem{p38}
\BIBentryALTinterwordspacing
M.~M. WHATWG~(Apple, Google, ``Html standard,'' 2021, (Accessed 30.10.2021).
  [Online]. Available: \url{https://html.spec.whatwg.org/}
\BIBentrySTDinterwordspacing

\bibitem{p6}
\BIBentryALTinterwordspacing
E.~International, ``Javascript,'' 2021, (Accessed 30.10.2021). [Online].
  Available:
  \url{https://www.ecma-international.org/publications-and-standards/standards/ecma-262/}
\BIBentrySTDinterwordspacing

\bibitem{p9}
R.~T. Fielding, \emph{Architectural styles and the design of network-based
  software architectures}.\hskip 1em plus 0.5em minus 0.4em\relax University of
  California, Irvine, 2000.

\bibitem{p22}
\BIBentryALTinterwordspacing
O.~Foundation and jQuery contributors, ``Ajax,'' 2021, (Accessed 30.10.2021).
  [Online]. Available: \url{https://api.jquery.com/}
\BIBentrySTDinterwordspacing

\bibitem{p11}
\BIBentryALTinterwordspacing
Google, ``Google maps,'' 2021, (Accessed 30.10.2021). [Online]. Available:
  \url{https://developers.google.com/maps}
\BIBentrySTDinterwordspacing

\bibitem{p36}
\BIBentryALTinterwordspacing
V.~Agafonkin, ``Leaflet,'' 2021, (Accessed 30.10.2021). [Online]. Available:
  \url{https://leafletjs.com/}
\BIBentrySTDinterwordspacing

\bibitem{p15}
\BIBentryALTinterwordspacing
JSON.org, ``Json,'' 2002, (Accessed 30.10.2021). [Online]. Available:
  \url{https:///www.json.org}
\BIBentrySTDinterwordspacing

\bibitem{p4}
\BIBentryALTinterwordspacing
{Chart.js contributors}, ``Chart.js,'' 2021, (Accessed 30.10.2021). [Online].
  Available: \url{https://www.chartjs.org/}
\BIBentrySTDinterwordspacing

\bibitem{p26}
\BIBentryALTinterwordspacing
Plotly, ``Plotly.js,'' 2021, (Accessed 30.10.2021). [Online]. Available:
  \url{https://plotly.com/}
\BIBentrySTDinterwordspacing

\bibitem{p8}
\BIBentryALTinterwordspacing
ESRI., ``Arcgis,'' 2021, (Accessed 30.10.2021). [Online]. Available:
  \url{https://www.esri.com/en-us/arcgis/products/arcgis-desktop/resources}
\BIBentrySTDinterwordspacing

\bibitem{p21}
\BIBentryALTinterwordspacing
O.~S.~G. Foundation., ``Geoserver.'' 2021, (Accessed 30.10.2021). [Online].
  Available: \url{http://geoserver.org/}
\BIBentrySTDinterwordspacing

\bibitem{p23}
\BIBentryALTinterwordspacing
O.~C. and/or~its affiliates, ``Mysql,'' 2021, (Accessed 30.10.2021). [Online].
  Available: \url{https://www.mysql.com/}
\BIBentrySTDinterwordspacing

\bibitem{p35}
\BIBentryALTinterwordspacing
T.~P. G.~D. Group, ``Postgresql.'' 2021, (Accessed 30.10.2021). [Online].
  Available: \url{https://www.postgresql.org}
\BIBentrySTDinterwordspacing

\bibitem{p24}
\BIBentryALTinterwordspacing
O.~C. and/or~its affiliates, ``Mysql spatial data types,'' 2021, (Accessed
  30.10.2021). [Online]. Available:
  \url{https://dev.mysql.com/doc/refman/8.0/en/spatial-types.html}
\BIBentrySTDinterwordspacing

\bibitem{p27}
\BIBentryALTinterwordspacing
P.~P.~S. Committee, ``Postgis,'' 2021, (Accessed 30.10.2021). [Online].
  Available: \url{https://postgis.net/}
\BIBentrySTDinterwordspacing

\bibitem{p2}
\BIBentryALTinterwordspacing
XXXXXXXXXXXXXX, ``Postgis: Data management and queries,'' (Accessed
  30.10.2021). [Online]. Available:
  \url{https://postgis.net/docs/manual-1.4/ch04.html}
\BIBentrySTDinterwordspacing

\bibitem{p28}
\BIBentryALTinterwordspacing
P.~P.~S. Committee, ``Postgis: Building applications,'' 2021, (Accessed
  30.10.2021). [Online]. Available:
  \url{https://postgis.net/docs/manual-1.4/ch05.html}
\BIBentrySTDinterwordspacing

\bibitem{p1}
\BIBentryALTinterwordspacing
XXXXXXXXXXXXXX, ``Object–relational mapping(orm),'' (Accessed 30.10.2021).
  [Online]. Available:
  \url{https://en.wikipedia.org/wiki/Object\%E2\%80\%93relational_mapping}
\BIBentrySTDinterwordspacing

\bibitem{p33}
\BIBentryALTinterwordspacing
S.~authors and contributors, ``Sqlalchemy.'' 2021, (Accessed 30.10.2021).
  [Online]. Available: \url{https://www.sqlalchemy.org}
\BIBentrySTDinterwordspacing

\bibitem{p5}
\BIBentryALTinterwordspacing
D.~S. Foundation, ``Django.'' 2021, (Accessed 30.10.2021). [Online]. Available:
  \url{https://www.djangoproject.com/}
\BIBentrySTDinterwordspacing

\bibitem{p34}
\BIBentryALTinterwordspacing
S.~authors and contributors, ``Sql inheritance hierarchies,'' 2021, (Accessed
  30.10.2021). [Online]. Available:
  \url{https://docs.sqlalchemy.org/en/14/orm/inheritance.html}
\BIBentrySTDinterwordspacing

\bibitem{p3}
\BIBentryALTinterwordspacing
Y.~Shafranovich, ``Comma-separated values,'' 2005, (Accessed 30.10.2021).
  [Online]. Available: \url{https://datatracker.ietf.org/doc/html/rfc4180}
\BIBentrySTDinterwordspacing

\bibitem{p29}
\BIBentryALTinterwordspacing
P.~S. Foundation, ``Python,'' 2021, (Accessed 30.10.2021). [Online]. Available:
  \url{https://www.python.org/}
\BIBentrySTDinterwordspacing

\bibitem{p25}
\BIBentryALTinterwordspacing
Pallets, ``Flask,'' 2010, (Accessed 30.10.2021). [Online]. Available:
  \url{https://flask.palletsprojects.com/en/2.0.x/}
\BIBentrySTDinterwordspacing

\bibitem{p7}
\BIBentryALTinterwordspacing
E.~Lemoine, ``Geoalchemy,'' 2012, (Accessed 30.10.2021). [Online]. Available:
  \url{https://www.geoalchemy-2.readthedocs.io/}
\BIBentrySTDinterwordspacing

\bibitem{p14}
\BIBentryALTinterwordspacing
ISO/IEC, ``Sql.'' 2016, (Accessed 30.10.2021). [Online]. Available:
  \url{https://www.iso.org/standard/63555.html/}
\BIBentrySTDinterwordspacing

\end{thebibliography}

\end{document}